# Transverse spin and transverse momentum in scattering of plane waves

SUDIPTA SAHA,[1,#] ANKIT K. SINGH,[1,#] SUBIR K. RAY,[1] AYAN BANERJEE,[1] SUBHASISH DUTTA GUPTA,[2] NIRMALYA GHOSH [1,*]

[1]*Department of Physical Sciences, Indian Institute of Science Education and Research-Kolkata, Mohanpur 741 246, India*
[2] *School of Physics, University of Hyderabad, Hyderabad-500046, India*
[#] *These authors contributed equally to this work.*
*\* Corresponding author: nghosh@iiserkol.ac.in*

***Abstract:-*** We study the near field to the far field evolution of spin angular momentum (SAM) density and the Poynting vector of the scattered waves from spherical scatterers. The results show that at the near field, the SAM density and the Poynting vector are dominated by their transverse components. While the former (transverse SAM) is independent of the helicity of the incident circular polarization state, the latter (transverse Poynting vector) depends upon the polarization state. It is further demonstrated that the magnitudes and the spatial extent of the transverse SAM and the transverse momentum components can be controllably enhanced by exploiting the interference of the transverse electric and transverse magnetic scattering modes.

It is well known that light wave carries both linear and angular momentum (AM) [1,2]. The plane wave or paraxial Gaussian beam typically carries longitudinal momentum and longitudinal spin AM (SAM, degree of circular/elliptical polarization or helicity,-1 ≤σ≤ +1). In addition to SAM, higher order Gaussian beams can also carry orbital AM [3]. The momentum and the two kinds of the AM of light play crucial roles in various light-matter interactions [4,5], and these are manifested as radiation-pressure force and torque (respectively) experienced by probe particles [6]. Recently it has been recognized that in addition to the conventional longitudinal AM (along the propagation direction specified by the wave vector ***k***), structured (inhomogeneous) optical fields can exhibit an unusual transverse (perpendicular to ***k***) SAM, which is independent of the helicity [6-8]. Moreover, such inhomogeneous fields also demonstrate polarization-dependent transverse momentum[9]. These unusual features are typically observed in the evanescent fields (e.g., for surface plasmon-polaritons at dielectric-metal interfaces) and leads to the optical spin-momentum locking in surface optical modes [5-7,10-12], similar to that observed for electrons in topological insulators.

The time-averaged densities of momentum ***P***, and SAM ***S*** are defined by the electric (***E***) and the magnetic fields (***H***) of light [13,14]. The momentum ***P*** is usually the canonical (orbital) momentum, responsible for the radiation pressure [4,6]. The transverse spin-momentum (***P_s*** ≈ ∇ × ***S***), introduced by Belinfante [15], on the other hand, is a virtual entity, which neither transfers energy nor exerts pressure on dipolar particles. The Poynting vector **P** contains contributions of both the orbital and the spin momentum [6]. For plane waves, **P** has longitudinal component solely due to the orbital momentum ***P***. For structured fields, on the other hand, finite longitudinal components of the fields are the sources of the transverse spin momentum $\mathcal{P}_s$. Moreover, a transverse component of SAM appears when the longitudinal field is shifted in phase with respect to the transverse field components. Finding out generalized systems and novel means to observe and amplify these entities is important from both fundamental context and potential spin-optical applications. Note that the general solution of scattering of plane waves from spherical particles yields a phase-shifted longitudinal field component (which is prominent in the near field) [14]. The question therefore arises – does scattering of plane waves produce such highly nontrivial structure of the momentum and spin densities? We show that indeed scattering leads to the generation of helicity-independent transverse SAM and polarization-dependent transverse momentum components. We study the near field to the far field evolution of the SAM density and the Poynting vector of the scattered waves. In order to explore ways to enhance the transverse SAM and transverse momentum components in scattering, we further analyze the influence of the localized plasmon resonances in metal nanospheres and the higher order transverse electric (TE) and transverse magnetic (TM) modes of dielectric microspheres on these quantities.

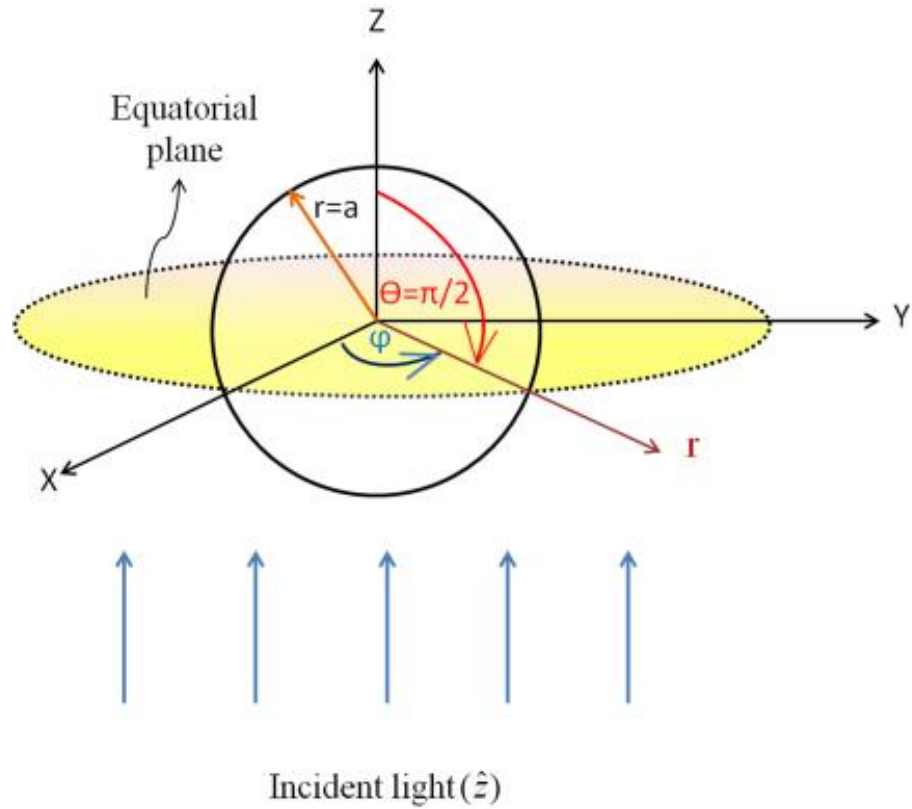

Fig.1. (Color online): Schematics of the scattering geometry. The SAM density and the Poynting vector distribution of the scattered wave are studied in the equatorial plane (scattering angle $\theta = \pi/2$) surrounding the spherical scatterer.

Consider the Cartesian coordinate system with the incident plane wave propagating in the $Z$ direction, the two orthogonal axes $X$ and $Y$ representing the polarization axes in the laboratory frame (**Figure 1**). The polar scattering angle $\theta$ and the azimuthal angle $\phi$ are assigned with respect to the Z and the X-axes, respectively. The expressions for the scattered electric ($E_s$) and the magnetic fields ($H_s$) for incident X-polarized plane wave can be obtained using Mie theory [14] as

$$E_s = \sum_{n=1}^{\infty} E_n(ia_n N^{(3)}_{e1n} - b_n M^{(3)}_{o1n})$$

$$H_s = \frac{k}{\omega\mu} \sum_{n=1}^{\infty} E_n(ib_n N^{(3)}_{o1n} + a_n M^{(3)}_{e1n}) \quad (1)$$

where $E_0$ is related to the amplitude of the incident plane wave, $a_n$ and $b_n$ are the Mie coefficients of the TM (electric) and TE (magnetic) scattering modes, $M$ and $N$ are the vector spherical harmonics [14]. In general, the scattered fields have all three components - the radial (longitudinal) component (specified by unit vector $\hat{r}$); the angular component ($\hat{\theta}$), perpendicular to the propagation direction but lying in the plane of scattering, the azimuthal component ($\hat{\varphi}$), perpendicular to both the propagation direction and the scattering plane. The expressions for the scattered fields for incident Y-polarized and for incident left (LCP) / right (RCP) circularly polarized light can be obtained using the symmetry of the scattering problem. These expressions for the scattered fields can then be utilized to obtain the normalized SAM density ($S$) and the Poynting vector ($P$) of the scattered wave as

$$S = \frac{\mathrm{Im}(\varepsilon E^* \times E + \mu H^* \times H)}{\omega(\varepsilon|E|^2 + \mu|H|^2)} \quad (2)$$

$$P = \frac{1}{2}\mathrm{Re}(E \times H^*) \quad (3)$$

Here, ($\varepsilon, \mu$) are the permittivity and permeability of the surrounding medium.

Using the expressions for the scattered fields (Eq. 1) in Eq. (2) and (3), it can be shown that irrespective of the scattering system, for incident LCP/RCP wave, the three components of $S$ and $P$ obey the following relations:

$$(S_r)_{LCP} = -(S_r)_{RCP}; (P_r)_{LCP} = (P_r)_{RCP}$$
$$(S_\theta)_{LCP} = -(S_\theta)_{RCP}; (P_\theta)_{LCP} = (P_\theta)_{RCP}$$
$$(S_\varphi)_{LCP} = (S_\varphi)_{RCP}; (P_\varphi)_{LCP} = -(P_\varphi)_{RCP} \quad (4)$$

The radial components ($S_r, P_r$) represent the longitudinal SAM density and longitudinal momentum, satisfying the usual dependence on the input wave helicity (SAM is opposite for $\sigma = \pm 1$, momentum is independent of $\sigma$). The angular components ($S_\theta, P_\theta$) also show this usual behavior. In contrast, for the azimuthal components, while $S_\varphi$ is independent of the helicity, $P_\varphi$ depends upon the input circular polarization state. Accordingly, $S_\varphi$ and $P_\varphi$ can be recognized as the helicity-independent transverse SAM and polarization-dependent transverse (spin) momentum, respectively.

Note that the SAM density (Eq. 2) has both electric and magnetic contributions ($S = S^e + S^m$), which are equivalent for plane (or paraxial) waves ($S^e \sim S^m$). However, for the scattered wave, the two may differ significantly because the longitudinal component of the scattered field can either be purely electric (for resonant TM- $a_n$ modes) or magnetic (for resonant TE- $b_n$ modes). Accordingly, $S_\varphi$ would either have electric or magnetic contributions. For non-resonant scatterers (having contributions of both the TM and TE modes), $S_\varphi$ would have both electric and magnetic contributions. It is also worth mentioning that the $P_\theta$ and the $P_\varphi$ Poynting vector components are the origin of the in-plane shift (eigen-modes- p and s-linear polarizations) [16] and the out of plane Spin-Hall shift (eigen-modes- LCP/RCP) of light in scattering [17]. Although, the exact expressions for $S$ and $P$ of the scattered wave are rather complicated (not presented here), it is not hard to recognize (using Eq. 1 in Eq. 2 and 3) that the components of $S$ and $P$ contain contribution of the interference of the modes (e.g., $Re\,[a_2^* a_1]$, when only the first and second order TM modes contribute). We therefore investigate the influence of the interference of the modes on the spin and momentum components for resonant localized plasmon modes in metal nanospheres (radius a<<λ), and for higher order TM and TE modes in non-resonant dielectric microspheres (a≥λ). For the ease of interpretation, we study the SAM density and the Poynting vector in the equatorial plane surrounding the spherical scatterer ($\theta = \pi/2$ in Figure 1).

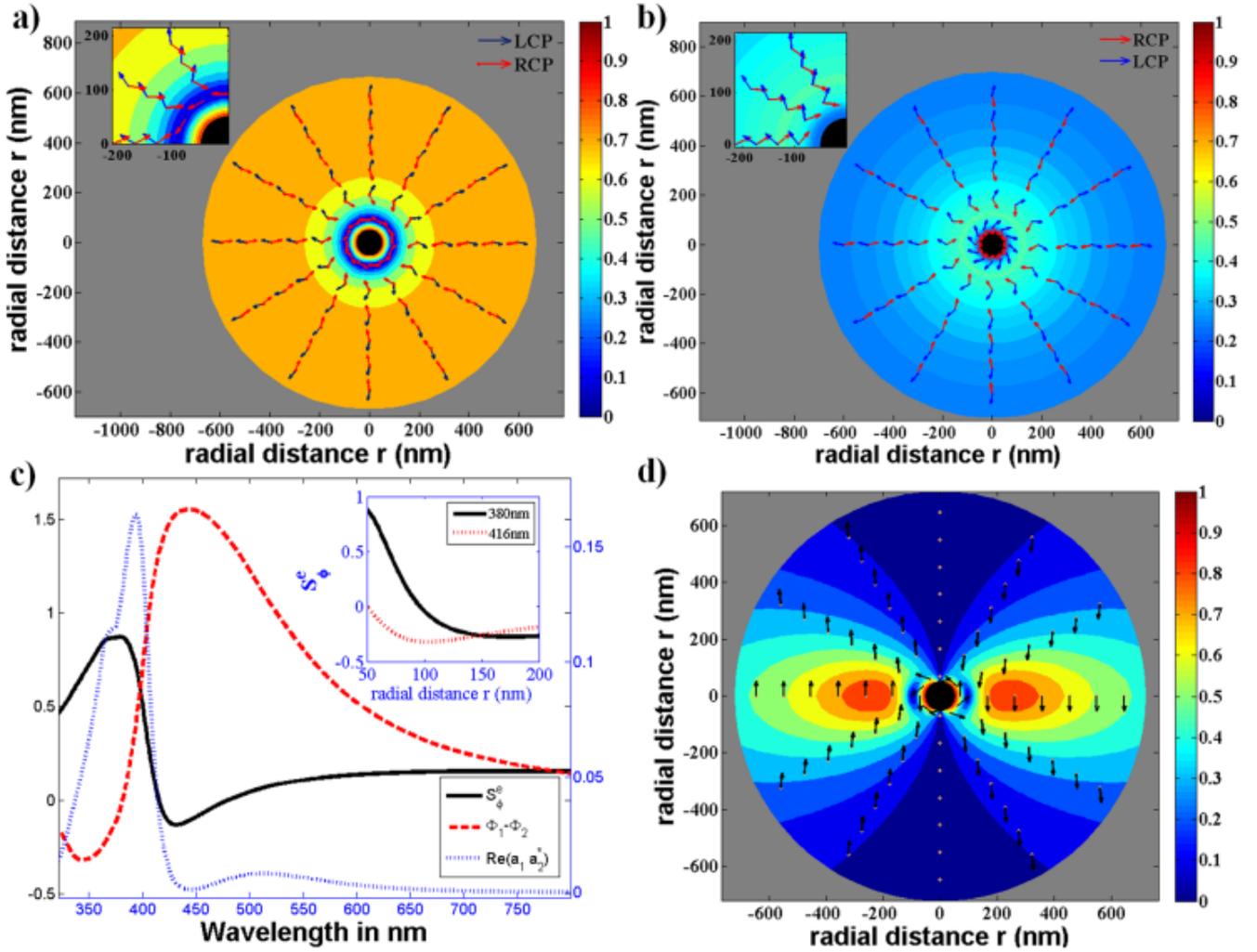

Fig. 2. (Color online): The computed electric part of the SAM density ($S^e$ shown in quiver plot) of the scattered wave from an Ag nanosphere ($a = 50$ nm) for incident LCP/RCP light, for (a) $\lambda = 380$ nm and (b) $\lambda = 416$ nm. (c) The wavelength dependence of the strength of the interference of dipolar $a_1$ and the quadrupolar $a_2$ plasmon modes ($\text{Re}(a_2^* a_1)$) (blue dotted line, right axis) and the phase difference ($\Phi$, in rad) between them (red dashed line, left axis). The wavelength dependence of the transverse SAM density ($S_\varphi^e$) at $r = 100$ nm (black solid line, left axis). The inset shows the comparison of the radial dependence of $S_\varphi^e$ at 380 nm (black solid line) and 416 nm (red dotted line). (d) Quiver plot of $S^e$ of the scattered wave at $\lambda = 380$ nm for incident X- polarized light. The magnitudes of $S^e$ are represented by colour code and the corresponding colour bars are displayed. The black circle represents the sphere at the equatorial plane here and in the subsequent figures.

**Figure 2** shows the computed SAM density ($\boldsymbol{S}$) of the scattered wave from a silver (Ag) nanosphere of radius $a = 50$ nm [18]. Here, the electric component of the SAM density ($\boldsymbol{S}^e$) is shown because for this metal nanosphere, scattering is primarily contributed by the TM-dipolar ($a_1$) and quadrupolar ($a_2$) plasmon modes (with their peaks at ~ 500 nm and 380 nm respectively) [16]. Accordingly, the transverse SAM $S_\varphi^e$ has only electric contributions. In the near field ($r \leq 150\,nm$), $\boldsymbol{S}^e$ is dominated by the transverse component ($S_\varphi^e$) (manifesting as azimuthal spin structure), which is independent of the input LCP/RCP state (Fig. 2a and 2b). As one moves away, the conventional longitudinal component ($S_r^e$) starts dominating, and in the far field ($r > 500\,nm$), the spin distribution exhibits radial structure and the expected behavior (opposite for input LCP/RCP states).

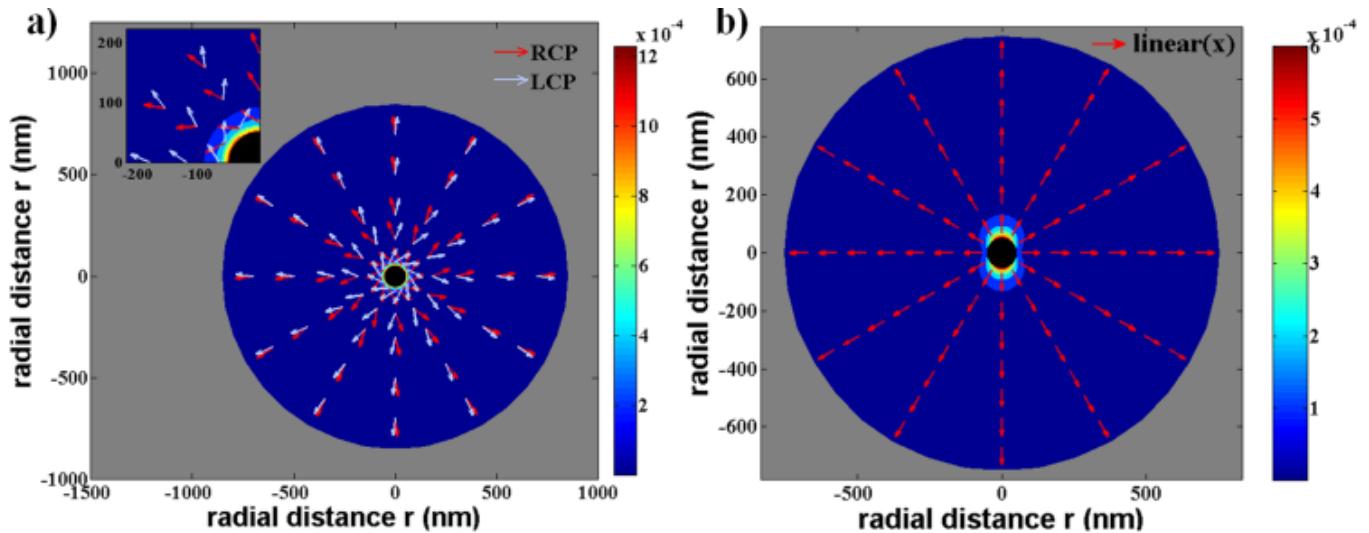

Fig.3. (Color online): The computed Poynting vector **P** distribution of the scattered wave (λ = 380 nm) from the Ag nanosphere corresponding to Fig. 2, for incident (a) LCP/RCP light and (b) X–linearly polarized light. The magnitudes of **P** are represented by colour code and the corresponding colour bars are displayed.

Note, the magnitude of the helicity-independent transverse SAM density at the near field is relatively stronger for λ = 380 nm (Fig. 2a) as compared to 416 nm (Fig. 2b). In order to understand this, in Fig. 2c, we show the wavelength dependence of the strength of the interference of the two modes ($Re(a_2^* a_1)$), and the phase difference ($\Phi = \phi_1 - \phi_2$) between them. The wavelength dependence of the transverse SAM density ($S_\varphi^e$) at a radial distance r= 100 nm is also shown. A comparison of the radial dependence of $S_\varphi^e$ at 380 nm and 416 nm is shown in the inset ($S_\varphi^e$ at 380 nm > $S_\varphi^e$ at 416 nm). As evident, $S_\varphi^e$ attains its maximum value at λ ~ 380 nm, where the strength of the interference is maximum (Φ~0° leading to constructive interference). The magnitude of $S_\varphi^e$ is minimum at λ ~ 420 nm, where the interference contribution is nearly vanishing (Φ~90°). Constructive interference of the two modes therefore leads to significant enhancement of the transverse spin in the near field. In Fig. 2d, the SAM density ($S^e$) at λ = 380 nm is displayed for incident X-polarized light. There are distinct regions in the near field where the transverse SAM density assumes significant magnitudes. Clearly, the phase shifted longitudinal component of the scattered electric field is responsible for this. Thus, the transverse SAM of the scattered wave in the near field is independent of the input helicity and also appears for incident linearly polarized light. This is therefore similar in nature to that observed for the evanescent waves [6].

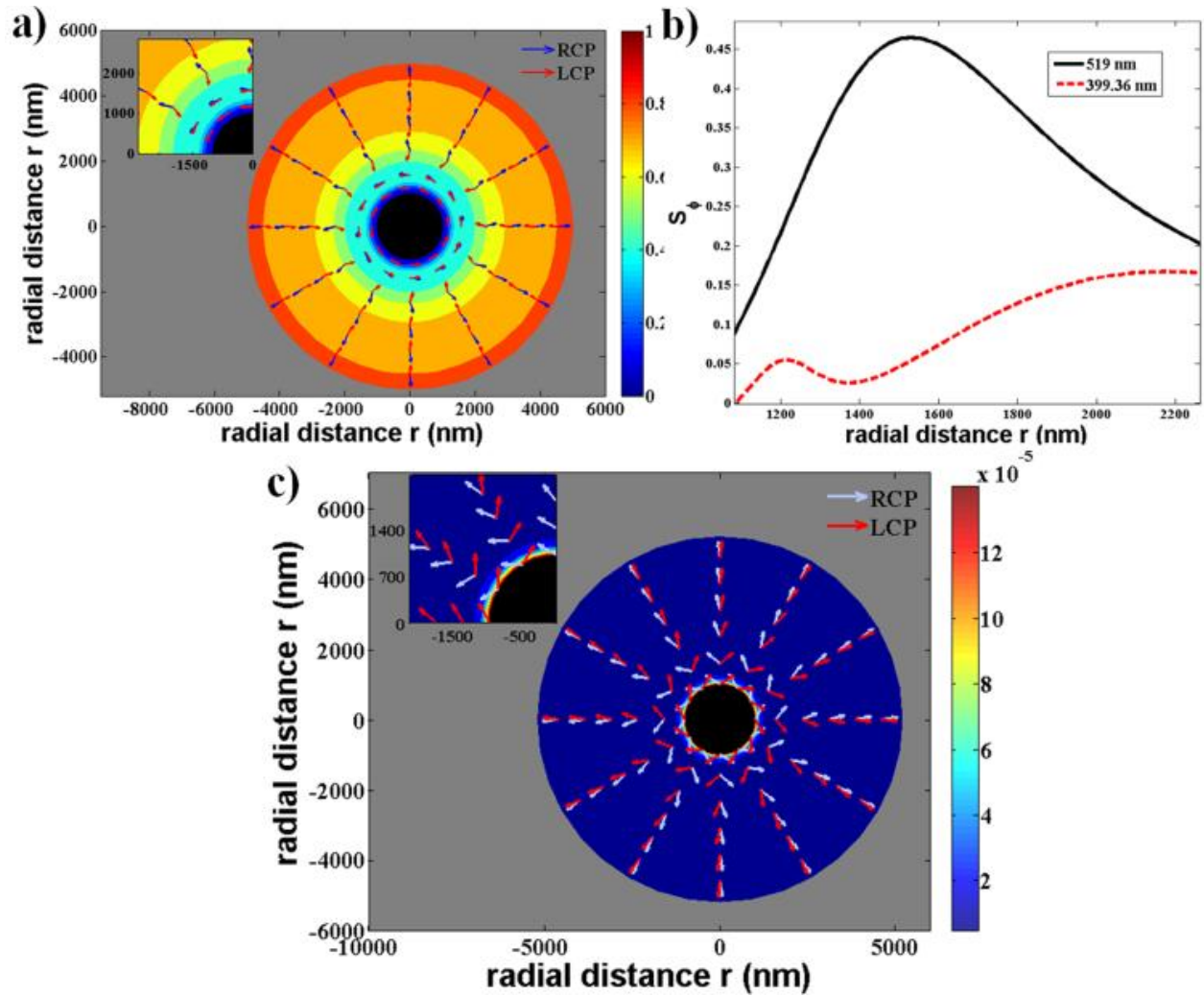

Fig.4. (Color online): The computed SAM density (**S**) and Poynting vector (**P**) distributions of the scattered wave from a dielectric microsphere ($a$ = 1μm and refractive index=1.6). (a) Quiver plot of **S** for λ = 519 nm. (b) Comparison of the radial dependence of transverse (azimuthal) SAM density ($S_\varphi$) at λ = 519 nm (non-resonant scattering system, black solid line) and λ = 399.36 nm (resonant scattering system, red dotted line). (c) Quiver plot of **P** of the scattered wave (λ = 519 nm) for incident LCP/RCP light.

In **Figure 3**, we show the Poynting vector (**P**) of the scattered wave ($\lambda$ = 380 nm) from the same Ag nanosphere for incident LCP/RCP (Fig. 3a) and X–polarized light (Fig. 3b). At the near field, **P** has significant contribution of transverse (azimuthal) component $P_\varphi$ which is opposite for input LCP/RCP states. With increasing distance, the conventional longitudinal (radial) component $P_r$ takes over and the usual polarization-independent radial energy flow is observed at the far field ($r > 500$ nm). Here also the magnitude of $P_\varphi$ was observed to be higher at $\lambda$ = 380 nm as compared to $\lambda$ = 416 nm (data not shown). The Poynting vector of the scattered wave for incident linearly polarized light (Fig. 3b), on the other hand, does not show any appreciable transverse component ($P_\varphi \sim 0$) and the energy flow is primarily in the radial direction. The circular polarization-dependent azimuthal Poynting vector component ($P_\varphi$) can therefore be interpreted as the transverse (spin) momentum. The results presented above demonstrate that the interference of the lower order TM modes (in plasmonic nanoparticle) leads to enhancement of the transverse SAM and transverse (spin) momentum components.

We now study how these quantities are affected when the higher order TM and TE ($a_n$ and $b_n$) modes are simultaneously excited. We choose a dielectric microsphere ($a = 1$ µm, refractive index =1.6) with air as surrounding medium. The SAM density (**S**) and Poynting vector (**P**) of the scattered wave from a dielectric scatterer (for $\lambda$ = 519 nm and 399.36 nm) are presented in **Figure 4**. At $\lambda$ = 519 nm, the dielectric microsphere acts as a non-resonant scatterer, where several higher order TM and TE modes contribute simultaneously. At $\lambda$ = 399.36 nm, on the other hand, scattering is primarily contributed by the resonant TM $a_{20}$ mode. The SAM density (**$S = S^e + S^m$**, Fig. 4a) at $\lambda$ = 519 nm exhibit similar trends as previously observed – (a) it is dominated by the helicity-independent transverse spin ($S_\varphi$) in the near field (exhibiting azimuthal structure); and (b) in the far field, it exhibits radial structure, which is opposite for incident LCP/RCP states. Importantly, the magnitude of $S_\varphi$ and its spatial extent is much larger for the non-resonant dielectric scatterer ($r \sim 2000\ nm$) as compared to that observed for the plasmonic Ag nanosphere ($r \sim 150\ nm$). Note that for this non-resonant scatterer, $S_\varphi$ has both electric and magnetic contributions ($S_\varphi^m \neq 0$). Even though, the resonant dielectric scattering system ($\lambda$ = 399.36 nm) also exhibits similar helicity-independent transverse SAM, its magnitude is weaker (Fig. 4b). The observed higher magnitude of $S_\varphi$ over an extended region for the non-resonant dielectric scattering system as compared to both the resonant dielectric scattering system and the plasmonic scatterer, provide conclusive evidence that simultaneous contribution of the higher order TM (electric) and TE (magnetic) modes and their resulting interference is responsible for the enhancement of the transverse spin. Finally, the Poynting vector (**P**) distribution of the scattered wave ($\lambda$ = 519 nm) (Fig. 4c), demonstrates input circular polarization-dependent transversal energy flow in the near field, with its spatial extent ($r \sim 1500\ nm$) being considerably larger as compared to that observed for the plasmon resonant Ag nanosphere ($r \sim 150\ nm$).

In conclusion, we have studied the near field to the far field evolution of spin angular momentum density and the Poynting vector components of the scattered waves from spherical scatterers. The results show dominant contribution of the helictiy-independent transverse spin and polarization-dependent transverse momentum components in the near field. It is demonstrated that the magnitude as well as the spatial extent of these unusual spin and momentum components can be controllably enhanced by exploiting the interference of the neighboring TM and TE scattering modes. The fact that the transverse

SAM and the transverse spin momentum can be observed and controllably enhanced in relatively simple optical systems, should prove valuable towards experimental observation and interpretation of these elusive fundamental entities. These findings should stimulate further experimental investigations on scattering from diverse micro/nano optical systems.

**Acknowledgements:-** The authors acknowledge IISER Kolkata for the funding and facilities. AKS and SKR acknowledge CSIR and UGC, Govt. of India (respectively) for research fellowships.